\documentclass[11pt,a4paper]{article}
\pdfoutput=1
\usepackage{empheq}
\usepackage{cases}
\usepackage{a4wide}
\usepackage{amsfonts}
\usepackage{amssymb}
\usepackage{amsmath,bm}
\usepackage{amsmath}
\usepackage{graphicx}
\usepackage{mathtools}
\usepackage{booktabs}
\usepackage{array}
\usepackage{color}
\usepackage{ulem}
\usepackage[small,bf]{caption}
\usepackage{cite}
\usepackage[usenames,dvipsnames]{xcolor}
\usepackage{subfigure}
\usepackage{verbatim}

\allowdisplaybreaks

\numberwithin{equation}{section}

\newcounter{mysubequation}[equation]

\DeclarePairedDelimiter\ket{\lvert}{\rangle}
\DeclarePairedDelimiterX\braket[2]{\langle}{\rangle}{#1 \delimsize\vert #2}

\begin{document}
\begin{titlepage}

\begin{center}
{
\bf\LARGE Probing Terrestrial Relic Neutrino Charge\\[0.3em]
with Mach-Zehnder Interferometer
}
\\[8mm]
Chuan-Ren~Chen\footnote[1]{crchen@ntnu.edu.tw},~Chrisna~Setyo~Nugroho\footnote[2]{setyo13nugros@ntnu.edu.tw},~and Vincent~Gene~L.~Otero\footnote[3]{vglotero@gmail.com}  
\\[1mm]
\end{center}
\vspace*{0.50cm}

\centerline{\it Department of Physics, National Taiwan Normal University, Taipei 116, Taiwan}
\vspace*{1.20cm}

\begin{abstract}
\noindent
We propose a novel method to probe the cosmic neutrino background (CNB) which has been shown to be accumulated 
on the surface of the earth. If such relic neutrino carries non-zero electric charge, Mach-Zehnder
interferometer offers a suitable venue to unveil  its interaction with photons. For neutrino mass equals to 0.05 eV, the sensitivity reach of our proposal could probe the fractional electric charge of the neutrino $\epsilon_{\nu}$ as low as $9.3 \times 10^{-11},\, 1.6 \times 10^{-16}$, and $2.9 \times 10^{-22}$ provided that the interferometer operates at standard quantum limit (SQL), the Heisenberg limit as well as super-Heisenberg limit, respectively. 
\end{abstract}

\end{titlepage}
\setcounter{footnote}{0}

\section{Introduction}

As an elusive particle within the standard model (SM) of particle physics, neutrino gives a direct hint
towards the existence of beyond standard model (BSM) thanks to its non-zero mass. Furthermore, other
unkown properties of the neutrino are still under active investigation.  
For example, it may have a non-zero electromagnetic moments~\cite{Ge:2022cib} and a non-zero electric
charge even within the framework of the SM~\cite{Foot:1990uf,Foot:1992ui}.
In particular, a minuscule electric charge of the neutrino would imply its Dirac mass origin and could
further reveal the nature of electric charge quantization~\cite{Babu:1989ex,Babu:1989wn}.
The upper limits on neutrino charge have been obtained from both laboratory experiments as well as astrophysical observations.

From cosmological frontier, the standard model of cosmology requires the existence of the relic
neutrino known as the cosmic neutrino background (CNB) produced in the early stage of the universe.  The detection of such relic
would allow us to gain a deeper understanding of the early cosmology as well as the nature of the neutrino itself. Unfortunately, this is not an easy task due to the smallness of the CNB energies and its respective cross sections with matter.
Several proposals have been made to discover the CNB such as utilizing the existing gravitational wave experiments~\cite{Domcke:2017aqj,Shergold:2021evs}, CNB absorption by tritium
in PTOLEMY experiment~\cite{PTOLEMY:2018jst,Betts:2013uya,Banerjee:2023lrk}, the enhanced
scattering between
ultra high energy cosmic neutrinos with the CNB~\cite{Brdar:2022kpu}, the induced cosmic
birefringence from CNB~\cite{Mohammadi:2021xoh}, and the
bremsstrahlung effect induced by the
CNB~\cite{Asteriadis:2022zmo}.
 
Recently, it has been shown that the scattering of relic neutrino with the earth leads to the
terrestrial accumulation of the CNB on the earth surface \cite{Arvanitaki:2022oby}. The study shows
that the number density of relic neutrino with respect to its anti-neutrino normalized to the CNB density in the universe $|n_{\nu} - n_{\bar{\nu}}|/n_\text{CNB}$ is of the order of $10^{-4}$. The corresponding asymmetry vanishes as one moves several meters away from the surface of the earth.  
If neutrino carries non-zero electric charge, it is possible to detect such terrestrial neutrino relic by utilizing laser interferometer.
The interaction between charged neutrino and the photon would induce a non-zero phase shift to be detected at the output port of the interferometer.    
However, it should be noted that the avaliable exotic particle search at Gravitational Wave (GW) experiments~\cite{Tsuchida:2019hhc,Lee:2020dcd,Chen:2021apc,Ismail:2022ukp,Lee:2022tsw,Seto:2004zu,Adams:2004pk,Nugroho:2024ltb,Riedel:2012ur,PhysRevLett.114.161301,Arvanitaki:2015iga,Stadnik:2015xbn,Branca:2016rez,Riedel:2016acj,Hall:2016usm,Jung:2017flg,Pierce:2018xmy,Morisaki:2018htj,Grote:2019uvn} can not be applied here. This is because the interferometer employed at GW experiments is either located horizontally on the ground or underground. Consequently, both arms of the interferometer would experience the same amount of relic neutrino leading to zero phase shift on the photons.

To tackle this issue, we propose to probe the electric charge of the terrestrial relic neutrino using
Mach-Zehnder (MZ) interferometer which has two perpendicular arms. One arm is placed horizontally on
the surface of the earth while another arm is positioned vertically underground. This configuration
allows us to probe non-zero phase shift induced by the relic neutrino since two arms of the
interferometer would face different neutrino density. We show that the projected sensitivities of our
proposal are quite competitive with respect to the existing bounds from laboratory experiments as well
as astrophysical observations.
 
This paper is structured as follows: We briefly discuss the terrestrial CNB accumulation in 
Section \ref{sec:Tcnb}. We further investigate the interaction between the non-relativistic charged neutrino and the photon in
Section~\ref{sec:interaction}. We introduce a phase measurement scheme 
based on Mach-Zehnder
interferometer in Section~\ref{sec:phase} and further show the corresponding
projected sensitivities of our proposal 
in section~\ref{sec:result}.
Finally, the summary and conclusion of our study are presented in Section~\ref{sec:Summary}.

\section{A Brief Review of Terrestrial CNB Accumulation}
\label{sec:Tcnb}

The accumulation of cosmic neutrino background on the surface of the earth is attributed to its
coherent interaction with ordinary matter via the weak force. This interaction causes some refractive
effects that alter the neutrino wave propagation near the surface, giving rise to a local $\nu-\bar{\nu}$ asymmetry. In general, neutrinos traversing ordinary matter experience in-matter potential $U$, derived from the weak 4-Fermi interactions \cite{Arvanitaki:2022oby}
\begin{equation}
\label{eq:Ucnb}
U = \frac{\sqrt{2}}{4} G_{F} \cdot Q_{W} \cdot \rho_{\text{matter}}\,,
\end{equation}
where $G_{F}$ is Fermi's constant, $\rho_{\text{matter}}$ is the material's atomic number density, and $Q_{W}$ is the weak charge of an atom for a given neutrino flavor. In particular, electron neutrinos
and muon or tau anti-neutrinos for which $Q_{W}>0$ and $U>0$, encounter a repulsive potential inside the
dense medium. Conversely, their CP-conjugate partners `feel' an attractive potential $U<0$ as $Q_{W}<0$. This potential produces an index of refraction $n^{\text{ref}}_\nu$ that deviates from that of the vacuum by an amount $\eta_\nu$ which is expressed as
\begin{equation}
 - \Biggl\langle\frac{m_\nu U}{k^2_\nu}\Biggr\rangle  \equiv \eta_\nu= n^{\text{ref}}_\nu - 1\,.
\end{equation}
Here, $m_\nu$ and $k_\nu$ are the neutrino mass and its momentum, respectively. Ref. \cite{Arvanitaki:2022oby} argues that the earth acts as a matter potential that leads to local enhancement of $\nu-\bar{\nu}$ asymmetry. In their geometric toy model, neutrinos -- modeled as incoherent superposition of plane waves -- are incident on a planar boundary with refractive index $|\eta_\nu| = 2.50 \times 10^{-8} $. This boundary separates two regions of space: vacuum and the earth's interior. It is assumed
that cosmic neutrinos and anti-neutrinos are incident from the vacuum side far from the interface with
momentum $k_\nu$. Neutrinos with momentum components perpendicular to the surface of the earth below
the critical value $\sqrt{2m_\nu U}$ reflect, intersecting the incoming neutrinos in vacuum. As a
result, neutrinos pile-up in a certain region of space in the vacuum side and thus increase the number
density, $n_\nu$, by a factor of 2. It can be deduced that the reflected rays are responsible for the
terrestrial accumulation of cosmic neutrinos around the surface of the earth. Since anti-neutrinos
experience an attractive potential, they are transmitted further inside the material and no reflected
waves are observed. Hence, this creates a net neutrino overdensity in a shell around the earth known as
local neutrino-anti-neutrino asymmetry. 

The aforementioned case is derived under the assumption that the earth is flat. From this approximation, the fractional excess in the neutrino number density given by 
\begin{equation}
\label{eq:TerestrialCNB}
n^{\text{asy}}_{\nu}=\frac{n_\nu-n_{\bar{\nu}}}{n_\text{CNB}}
\end{equation} is determined to be of order $\mathcal{O}(10^{-4})$, which is several orders of magnitude larger than the expected primordial lepton asymmetry $\sim\mathcal{O}(10^{-9})$. However,
this calculation is oversimplified and does not account for the realistic geometry of the earth. On the
other hand, Ref. \cite{Kalia:2024xeq} revisited this analysis and extended the calculations using a
perfectly spherical earth configuration. It was shown that the flat earth approximation is valid only
under the condition $\eta^{3/2}_\nu k_\nu R_E>>1$, where $R_E$ is the earth's radius. In the proposed
spherical model, quantum tunnelling permits antineutrinos to penetrate the classically inaccessible
region beneath the surface, effectively cancelling out the asymmetric accumulation of neutrinos. The
analytic and numerical treatment found that the fractional asymmetry is of order $\mathcal{O}(10^{-8})$. This conclusion has been verified independently by \cite{Huang:2024tog} through a full solution scattering problem for the perfectly round Earth by employing partial wave expansion. The net asymmetry
was rigorously calculated by summing all the angular modes and found constrained to the maximal value
of order $\mathcal{O}(10^{-8})$ which corroborates the previous spherical model. This study has been
confirmed by using thermal method in \cite{Gruzinov:2024ciz} which also shows the overdensity is of the
order of $10^{-8}$.
In this paper, we set the value of the neutrino asymmetry in Eq. \eqref{eq:TerestrialCNB} to $10^{-8}$. In addition, we fix the neutrino number density in the universe $n_{\text{CNB}}$ equals to $56 \, \text{cm}^{-3}$~\cite{Ringwald:2004np}.

\section{CNB and Photon Interaction}
\label{sec:interaction}

The temperature of CNB today is about 1.95 K implying its non-relativistic nature as opposed to the other known neutrino sources. Thus, to probe the electric charge of the CNB in a laser interferometry setup, one needs to consider the
interaction between photon and non-relativistic charged particles which is given by the following Hamiltonian 
\begin{align}
\label{eq:hamiltontot}
H = H_{P} + H_{R} + H_{I} \,.
\end{align}
Here, $H_{P}$, $H_{R}$, and $H_{I}$ correspond to the  Hamiltonian of the charged
particles, non-intercting radiation field, and the interaction between radiation and non-relativistic charged particles, respectively.
 Their explicit expression is written as~\cite{cohen:1987}
\begin{align}
\label{eq:hamilall}
H_{P} &= \sum_{s} \frac{\vec{p}^{2}_{s}}{2\, m_{s}} + V_{\text{Coulomb}}\,,\\
H_{R} &= \sum_{i} \hbar \omega_{i} \left( \hat{a}^{\dagger}_{i} \hat{a}_{i} + \frac{1}{2}\right)\,,\\
H_{I} &= H_{I1} + H_{I2}\,,\\
H_{I1} &= - \sum_{s} \frac{\text{q}_{s}}{m_{s}} \, \vec{p}_{s} \cdot \vec{A}(\vec{r}_{s})\,,\\
H_{I2} &= \sum_{s} \frac{\text{q}^{2}_{s}}{2\,m_{s}} \left[ \vec{A}(\vec{r}_{s})\right]^{2}\,.
\end{align}  
In above equations, $\vec{p}_{s}$, $m_{s}$, and $\text{q}_{s}$
are the momentum, the mass, and the electric charge of the  $s$-th non-relativistic charged particle, respectively. Furthermore, we have quantized the radiation field with $\hat{a}_{i}$ ($\hat{a}^{\dagger}_{i}$) denoting the annihilation (creation)
operator of the photon field for the i-th mode with the known
commutation relation $[\hat{a}_{i},\hat{a}^{\dagger}_{j} ] = \delta_{ij}$. In term of these operators, the radiation field can be written as~\cite{cohen:1987}
\begin{align}
\label{eq:photon}
\vec{A}(\vec{r}) = \sum_{i} \left[ \frac{\hbar}{2 \,\epsilon_{0}\, \omega_{i} L^{3}} \right]^{1/2} \left[\hat{a}_{i}\, \vec{\varepsilon}_{i} \, e^{\text{i} \vec{k}_{i} \cdot \vec{r}} + \hat{a}^{\dagger}_{i}\, \vec{\varepsilon}_{i} \, e^{-\text{i} \vec{k}_{i} \cdot \vec{r}} \right] \,,
\end{align}
where we have employed the box quantization of the radiation field $\vec{A}(\vec{r})$ in a volume of $L^{3}$ with the corresponding boundary condition $\vec{k} \cdot \vec{L} = 2\pi\,n$ with $n$ equals to integer.
It is worth mentioning that the relation between the wave number and the angular frequency of the photon is given by $ \omega = |\vec{k}|\, c$. 

In a laser interferometry setup, the induced phase shift $\delta$ comes from the interaction between the photon and the charged particles. In other
words, the only relevant terms in the Hamiltonian are $H_{I} = H_{I1} + H_{I2}$. We could further drop $H_{I1}$ term since it would induced one photon absorption irrelevant to the unbound system considered here.
Moreover, not all terms in $H_{I2}  \propto (\hat{a} \hat{a} + \hat{a} \hat{a}^{\dagger} + \hat{a}^{\dagger} \hat{a} + \hat{a}^{\dagger} \hat{a}^{\dagger})$ contribute to the phase shift, since the first (last) term would induce two photon absorption (emission) which violates the energy conservation for unbound particle system. Therefore, one arrives at the following expression
\begin{align}
\label{eq:HintF}
H_{I} \equiv \hat{H}_{\text{int}} &= \sum_{s} \frac{q^{2}_{s}}{2\,m_{s}} \, \left[ \frac{\hbar}{2 \,\epsilon_{0}\, \omega_{i} L^{3}} \right] 2\left(\hat{a}^{\dagger} \hat{a} + \frac{1}{2} \right)\,,\nonumber \\
&= \frac{\epsilon^{2}_{\nu} \, e^{2}}{m_{\nu}}\,\left[ \frac{\hbar\,\omega^{2}}{16\,\pi^{3} \,\epsilon_{0}\, c^{3}} \right] \left(\hat{a}^{\dagger} \hat{a} + \frac{1}{2} \right)\, N_{\nu}\,. 
\end{align} 
Here, we set all CNB neutrinos to have equal electric charge $q_{s} = \epsilon_{\nu}\,e$ and the same mass $m_{s} = m_{\nu}$ such that
the sum over all neutrinos is proportional to the total number of CNB neutrinos interacting with photons $N_{\nu}$. Moreover, in the second line of Eq.\eqref{eq:HintF}, we have substituted $L = 2\pi/ k$ for $n = 1$ or single mode field analogous to the Jaynes-Cummings model in atomic physics~\cite{Jaynes:1963zz,Garrison:2008jh} which has been realized in the laboratory~\cite{Fox:2006quantum}.

\begin{figure}
	\centering
	\includegraphics[width=0.9\textwidth]{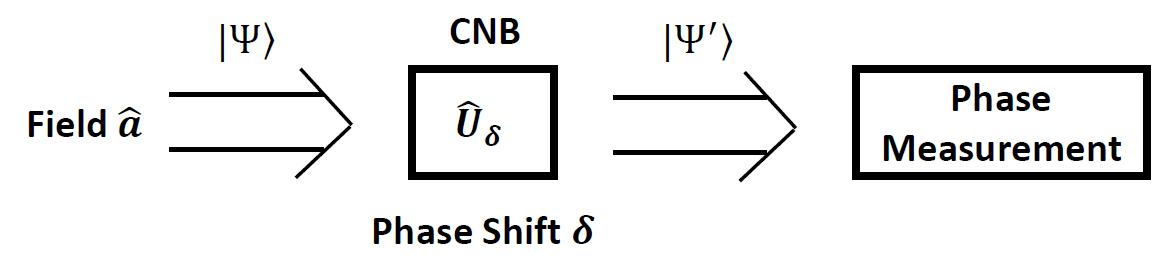}
	\caption{The induced phase shift $\delta$ from photon-CNB interaction would transform the photon state $\ket{\Psi}$ to $\ket{\Psi^{'}}$~\cite{Nugroho:2023cun}.}
	\label{fig:interfero}
\end{figure}
In quantum mechanics, the interaction between photon and CNB would transform the initial state of the photon $\ket{\Psi}$ into $\ket{\Psi^{'}}$. These two states are related via unitary operator
$\hat{U}_{\delta}$~\cite{Kartner:1993,Ou:2017}
\begin{align}
\label{eq:psiP}
\ket{\Psi^{'}} &= \hat{U}_{\delta} \, \ket{\Psi} =e^{-\text{i}\,\hat{H}_{\text{int}}\text{t}/\hbar}\, \ket{\Psi}= e^{-\text{i}\,\hat{N}\delta}\,\ket{\Psi}\,, 
\end{align}
where $ \hat{N} \equiv \hat{a}^{\dagger}\hat{a}$ is the photon number operator with the average value
$ N \equiv \left\langle \hat{a}^{\dagger}\hat{a} \right\rangle\gg 1$. To measure the induced phase shift, one needs to prepare a suitable phase measurement shown in Fig.\ref{fig:interfero}.   
Unfortunately,  the quantum nature of the light prohibits us to extract the phase shift as accurate as possible. The minimum detectable phase shift is given by the Heisenberg
limit~\cite{Dirac:1927, Heitler:1954}
\begin{align}
\label{eq:HeisLimit}
\Delta \delta \geq \frac{1}{N_{ps}}\,,
\end{align}
where $N_{ps}$ is the total number of photon utilized to probe the phase shift. In addition, there is the so called the standard quantum limit (SQL) of the phase measurement which corresponds to $\Delta \delta \geq 1/N^{1/2}_{ps}$. Typically,
the number of photon in modern laser interferometer is of the
order of $10^{20}$ or larger which enbles us to probe minuscule phase in the lab.

\section{CNB Induced Phase Measurement Scheme}
\label{sec:phase}
It is well known that laser interferometer has been widely used to measure the phase shift of the
photon which originates from its arm length difference. The state-of-the-art of modern laser
interferometer has reached its pinnacle at the discovery of gravitational wave (GW) by the LIGO
collaboration~\cite{LIGOScientific:2016aoc}. However, one can not employ the existing terrestrial GW
experiments to probe the CNB electric charge since both arms of the interferometers are located at the same depth.

\begin{figure}
	\centering
	\includegraphics[width=0.9\textwidth]{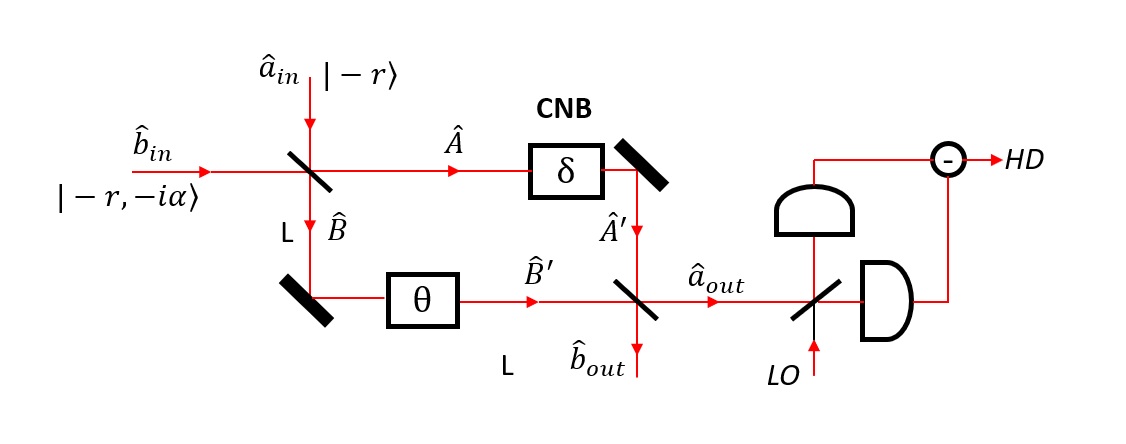}
	\caption{Mach-Zehnder interferometer  with two light sources at its input: the squeezed vacuum state and the squeezed coherent state. The homodyne detection (HD) method using local oscillator (LO) is utilized to measure the induced phase shift $\delta$ from CNB-photon interaction.}
	\label{fig:MZInterferometer}
\end{figure}

To overcome this issue, we propose to employ Mach-Zehnder interferometer shown in Fig.~\ref{fig:MZInterferometer}. This interferometer, which was originally invented by~\cite{Bondurant:1984}, has been shown to overcome the SQL along with other type of interferometers~\cite{Grangier:1987,Xiao:1987}. Moreover, Mach-Zehnder interferometer has been proposed to detect terrestrial millicharged particles with impressive sensitivity reach~\cite{Chen:2022abz}.

In Fig.\ref{fig:MZInterferometer}, we show the Mach-Zehnder (MZ) interferometer which consists of two perpendicular
arms. One arm with the photon field $\hat{A}$ is placed horizontally on the earth surface
while another arm corresponding to the photon field $\hat{B}$ is positioned vertically underground.
Furthermore, instead of using single laser source in its input port, we utilize squeezed vacuum state $\ket{-r}$ together with squeezed coherent
state $\ket{-r,-i\alpha}$~\cite{Bondurant:1984}. In this case, 
the squeezed vacuum state can be written in term of the squeezing operator $\hat{S}(r)$ with its associated real squeezing parameter $r$ as~\cite{Ou:2017}
\begin{align}
\label{eq:defsqueeze}
\ket{r} = e^{r (\hat{a}^{\dagger\,2} - \hat{a}^{2})/2} \ket{0} \equiv \hat{S}(r) \ket{0}\,,
\end{align}
where we have used used the photon number basis $\ket{n}$ with $\ket{0}$ describing zero photon number or the vacuum state.
In this basis,
the
coherent state $\ket{\alpha}$ is given by
\begin{align}
\label{eq:DefCoherent}
\ket{\alpha} = e^{-|\alpha|^{2}/2} \sum^{\infty}_{n = 0} \frac{\alpha^{n}}{\sqrt{n !}}\, \ket{n} \equiv \hat{D}(\alpha) \ket{0}\,. 
\end{align} 
Here, $\hat{D}(\alpha) = e^{\alpha\hat{a}^{\dagger}-\alpha^{*} \hat{a}}$ denotes the displacement operator with its corresponding complex displacement parameter $\alpha$.  
Using these operators, one can construct the squeezed coherent state  $\ket{r,\alpha}$ from the vacuum state
\begin{align}
\label{eq:SCoh}
\ket{r,\alpha} = \hat{S}(r)\,\hat{D}(\alpha)\,\ket{0}\,.
\end{align}  
Moroeover, the squeezing
operator and the displacement operator transform the photon field operators $\hat{a}$ and $\hat{a}^{\dagger}$ into the following
\begin{align}
\label{eq:OpProperties}
\hat{S}^{\dagger}(r)\, \hat{a}\, \hat{S}(r) &= \hat{a}\, \text{cosh}\,r \, + \hat{a}^{\dagger} \, \text{sinh}\,r\,, \\
\hat{S}^{\dagger}(r)\, \hat{a}^{\dagger}\, \hat{S}(r) &= \hat{a}^{\dagger}\, \text{cosh}\,r \, + \hat{a} \, \text{sinh}\,r\,, \\
\hat{D}^{\dagger}(\alpha)\, \hat{a}\,\hat{D}(\alpha) &= \hat{a} \, +\, \alpha\,, \\
\hat{D}^{\dagger}(\alpha)\, \hat{a}^{\dagger}\,\hat{D}(\alpha) &= \hat{a}^{\dagger} \, +\, \alpha^{*}\,.
\end{align}
In the subsequent discussion, we assume that the displacement parameter $\alpha$ has the real value.

As shown in Fig.\ref{fig:MZInterferometer}, two light sources correspond to the photon field $\hat{a}_{in}$ and $\hat{b}_{in}$ enter the first 50:50 beam splitter which
splits the photon path associated with two operators $\hat{A}$ and $\hat{B}$. Subsequently, due to the
interaction between the photon field $\hat{A}$ and the CNB, it transforms into $\hat{A}^{'}$. On the other
hand, one may calibrate the interferometer in such a way that the
operator $\hat{B}$ would carry an adjustable phase $\theta$ such that $\hat{B}^{'} = \hat{B} e^{\text{i} \theta}$. Next, two beams $\hat{A}^{'}$ and $\hat{B}^{'}$ recombine at the second beam splitter before reaching the output ports. Finally, two independent beams $\hat{a}_{out}$ and $\hat{b}_{out}$ reach two different output ports perpendicular to each other. The phase measurement is carried out on the photon field $\hat{a}_{out}$ (the dark port) using the local oscillator (LO) via homodyne detection (HD) method. Note that the phase adjustment on the photon field $\hat{B}^{'}$ is useful because in the absence of CNB-photon
interaction, one may set the dark fringe output located at $\hat{a}_{out}$ to give zero photon number.

Furthermore, since both input fields are independent of each other, one can set the operator $\hat{a}_{in}$ to act on the state $\ket{-r}$ while the operator $\hat{b}_{in}$ operates on $\ket{-r,-i\alpha}$. The explicit expression of the operators $\hat{A}$, $\hat{B}$, $\hat{A}^{'}$, $\hat{B}^{'}$, $\hat{a}_{out}$, and $\hat{b}_{out}$ shown in Fig.\ref{fig:MZInterferometer} are
\begin{align}
\label{eq:ABDefinition}
\hat{A} &= \frac{(\hat{a}_{in} + \hat{b}_{in})}{\sqrt{2}},\,\,\,\,\,\hat{B} = \frac{(-\hat{a}_{in} + \hat{b}_{in})}{\sqrt{2}}\,,\\
\hat{A}^{'} &= \hat{A}\,e^{\text{i}\delta},\,\,\,\,\,\hat{B}^{'} = \hat{B}\,e^{\text{i}\theta}\,,\\
\hat{a}_{out} &= \frac{(\hat{A}^{'} - \hat{B}^{'})}{\sqrt{2}},\,\,\,\,\,\hat{b}_{out} = \frac{(\hat{A}^{'} + \hat{B}^{'})}{\sqrt{2}}\,.
\end{align}
In this paper, we fix $\theta = \pi$ such that the dark fringe output $\hat{a}_{out}$ is given by
\begin{align}
\label{eq:aout}
\hat{a}_{out} = \text{i}\, e^{i\delta/2}\, \left(\hat{a}_{in}\,\text{sin}\,\frac{\delta}{2}-\text{i}\,\hat{b}_{in}\, \text{cos}\,\frac{\delta}{2} \right)\,.
\end{align}
Moreover, we measure the quadrature amplitude $\hat{X}_{a} = \hat{a}_{out} + \hat{a}^{\dagger}_{out}$ at the output port by using the homodyne detection (HD) method such that
\begin{align}
\label{eq:XaDelta}
\hat{X}_{a} = -\text{sin}\,\frac{\delta}{2}\,\hat{Y}_{a_{in}}(-\delta/2) + \text{cos}\,\frac{\delta}{2}\,\hat{X}_{b_{in}}(-\delta/2)\,.
\end{align}
Here, the two operators $\hat{Y}_{a_{in}}(-\delta/2)$ and $\hat{X}_{b_{in}}(-\delta/2)$ are defined as
\begin{align}
\label{eq:XY}
\hat{Y}_{a_{in}}(-\delta/2) &= -\text{i}\,\left(\hat{a}_{in}\,e^{\text{i}\,\delta/2} - \hat{a}^{\dagger}_{in}\,e^{-\text{i}\,\delta/2} \right)\,,\\
\hat{X}_{b_{in}}(-\delta/2) &= \left(\hat{b}_{in}\,e^{\text{i}\,\delta/2} + \hat{b}^{\dagger}_{in}\,e^{-\text{i}\,\delta/2} \right)\,.
\end{align}
In this phase measurement scheme, the sensitivity reach is obtained by evaluating the expectation value of the quadrature amplitude $\hat{X}_{a}$ as well as its corresponding fluctuation $\Delta^{2} \hat{X}_{a}$
\begin{align}
\label{eq:evXa}
\left\langle\hat{X}_{a}\right\rangle &= \alpha\,(\mu + \nu)\,\text{sin}\,\delta\,,\\
\left\langle\Delta^{2} \hat{X}_{a}\right\rangle &= \mu^{2} + \nu^{2} - 2\,\mu\, \nu\, \text{cos}^{2}\delta\,,
\end{align}
where $\mu = \text{cosh}\,r$ and $\nu = \text{sinh}\,r$. The corresponding signal-to-noise ratio (SNR) in this case is given by~\cite{Ou:2017}
\begin{align}
\label{eq:SNR}
\text{SNR} \equiv \frac{\left\langle\hat{X}_{a}\right\rangle^{2}}{\left\langle\Delta^{2} \hat{X}_{a}\right\rangle} = \frac{\alpha^{2}\,(\mu + \nu)^{2}\, \text{sin}^{2}\, \delta}{\mu^{2} + \nu^{2} - 2\,\mu\, \nu\, \text{cos}^{2}\delta}\,.
\end{align}

Furthermore, one can determine the phase sensing photon number as $N_{ps} \equiv \left\langle\hat{A}^{\dagger} \hat{A}\right\rangle = \nu^{2} + \alpha^{2} (\mu + \nu)^{2} /2$. In addition, we fix $N_{ps}$ to be a constant and set $\nu \gg 1$ in such a way that $\mu - \nu = 1/(\mu + \nu) \approx 1/2\nu$. Thus, the associated SNR becomes
\begin{align}
\label{eq:SNRfin}
\text{SNR} = 4\, (N_{ps} -\nu^{2})\, \nu^{2}\,\text{sin}^{2}\,\delta \leq  N^{2}_{ps}\,\text{sin}^{2}\,\delta\,.
\end{align}
Moreover, the maximum value of the SNR is reached when $\nu^{2} = N_{ps}/2$. Note that in the limit $\delta \ll 1$ and $\text{SNR} \sim 1$, the minimum detectable phase shift is $\delta_{min} \sim 1/N_{ps}$ or the Heisenberg limit. In the rest of the paper, we take the maximum value of the SNR in Eq.~\eqref{eq:SNRfin}.

Several remarks are in order regarding Eq.~\eqref{eq:SNRfin}. There, the noise component of the SNR comes from quantum
fluctuation of the photon. In addition, one can safely drop other noise components such as imperfect particle detection as well as photon loss since
they can be alleviated using current
technology~\cite{Szigeti:2017}. Experimentally, the Heisenberg limit has been realized in the
laboratories~\cite{Szigeti:2017,Linnemann:2016, Daryanoosh:2018, Anderson:2017}.
Moreover, the sensitivity surpassing the Heisenberg limit ($\delta \geq 1/N^{3/2}_{ps}$) called the "super-Heisenberg limit" has been carried out in quantum optic experimental setup by utilizing $\mathcal{O} (10^{7})$ photons \cite{Napolitano:2011, Napolitano:2011b}. In addition, recent study shows that it is possible to go beyond the super-Heisenberg limit by utilizing Mach-Zehnder interferometer to achieve the phase sensitivity $\delta \geq 1/N^{2}_{ps}$~\cite{Qin:2023ift}. In this paper, we consider three different operating modes of the Mach-Zender interferometer under consideration: the SQL, the Heisenberg limit, as well as the potential super-Heisenberg limit.
\begin{figure}
	\centering
	\subfigure{{\includegraphics[width=14.0cm]{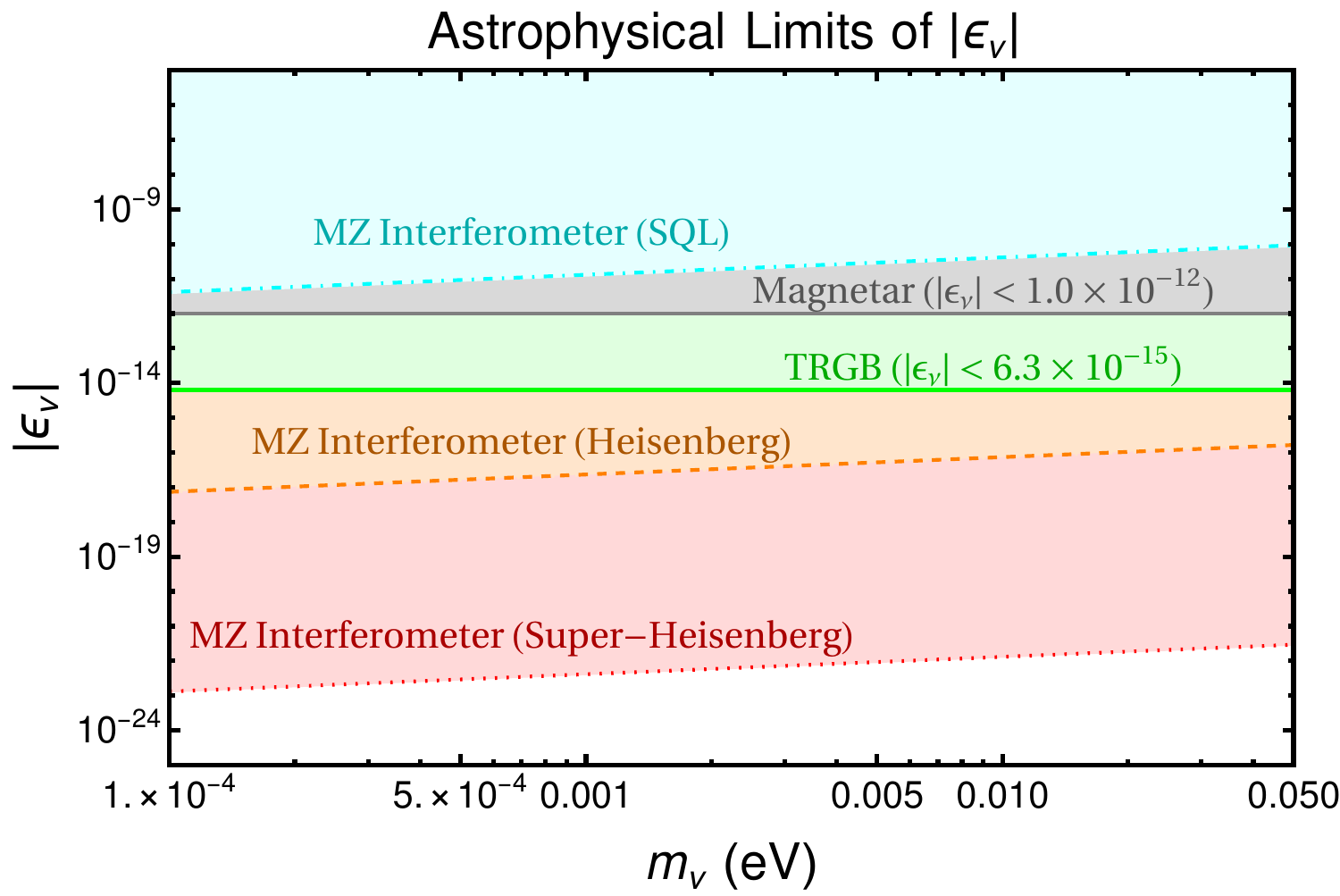} }}
	\caption{The sensitivity reach of the MZ interferometer with arm length L = 1 km and 1.17 eV laser for $|n_{\nu} - n_{\bar{\nu}}|/n_{\text{CNB}} = 10^{-8}$ vs the existing astrophysical bounds. Here, we take $n_{\text{CNB}} = 56\, \text{cm}^{-3}$ and set the phase sensing photon number $N_{ps} = 10^{23}$~\cite{GammeVT-969:2007pci,Bahre:2013ywa,ALPS:2009des,Inada:2013tx}. } 
	\label{fig:sensitivity_astro}
\end{figure}  

\section{Projected Sensitivity}
\label{sec:result}
Before discussing the projected sensitiviy of the MZ interferometer, one needs to extract the explicit expression of the phase shift $\delta$. This can be determined by comparing Eq.~\eqref{eq:HintF} and \eqref{eq:psiP} to get
\begin{align}
\label{eq:delta}
\delta = \frac{\epsilon^{2}_{\nu} \, e^{2}}{m_{\nu}}\,\left[ \frac{\omega^{2}}{16\,\pi^{3} \,\epsilon_{0}\, c^{3}} \right] \, N_{\nu}\,t\,,
\end{align}     
which explicitly shows the the phase shift as a function of the the total number $N_{\nu}$, fractional electric charge $\epsilon_{\nu}$, and the mass $m_{\nu}$ of the CNB neutrino.
The total number of CNB $N_{\nu}$ is determined by integrating the number of CNB per unit length which interacts with the phase sensing photon $\hat{A}$ along its path $\ell$
\begin{align}
\label{eq:ell}
N_{\nu} = \int^{\text{L}}_{0} d\ell\, \tilde{n}_{\nu}\,.
\end{align}
Here, $\tilde{n}_{\nu} = n^{1/3}_{\nu^{*}}$ and L stand for the number of asymmetric CNB per unit length in $\text{cm}^{-1}$ and the arm length of the interferometer, respectively. For asymmetric terrestrial CNB considered here, the appropriate number density is $n_{\nu^{*}} \equiv |n_{\nu} - n_{\bar{\nu}}| = n_{\text{CNB}} \times 10^{-8}\, \text{cm}^{-3}$ with the number density of the relic neutrino in the universe $n_{\text{CNB}} = 56\,\text{cm}^{-3}$. Finally, the projected sensitivity is obtained by requiring the SNR to be larger than one.

\begin{figure}
	\centering
    \subfigure{{\includegraphics[width=14.0cm]{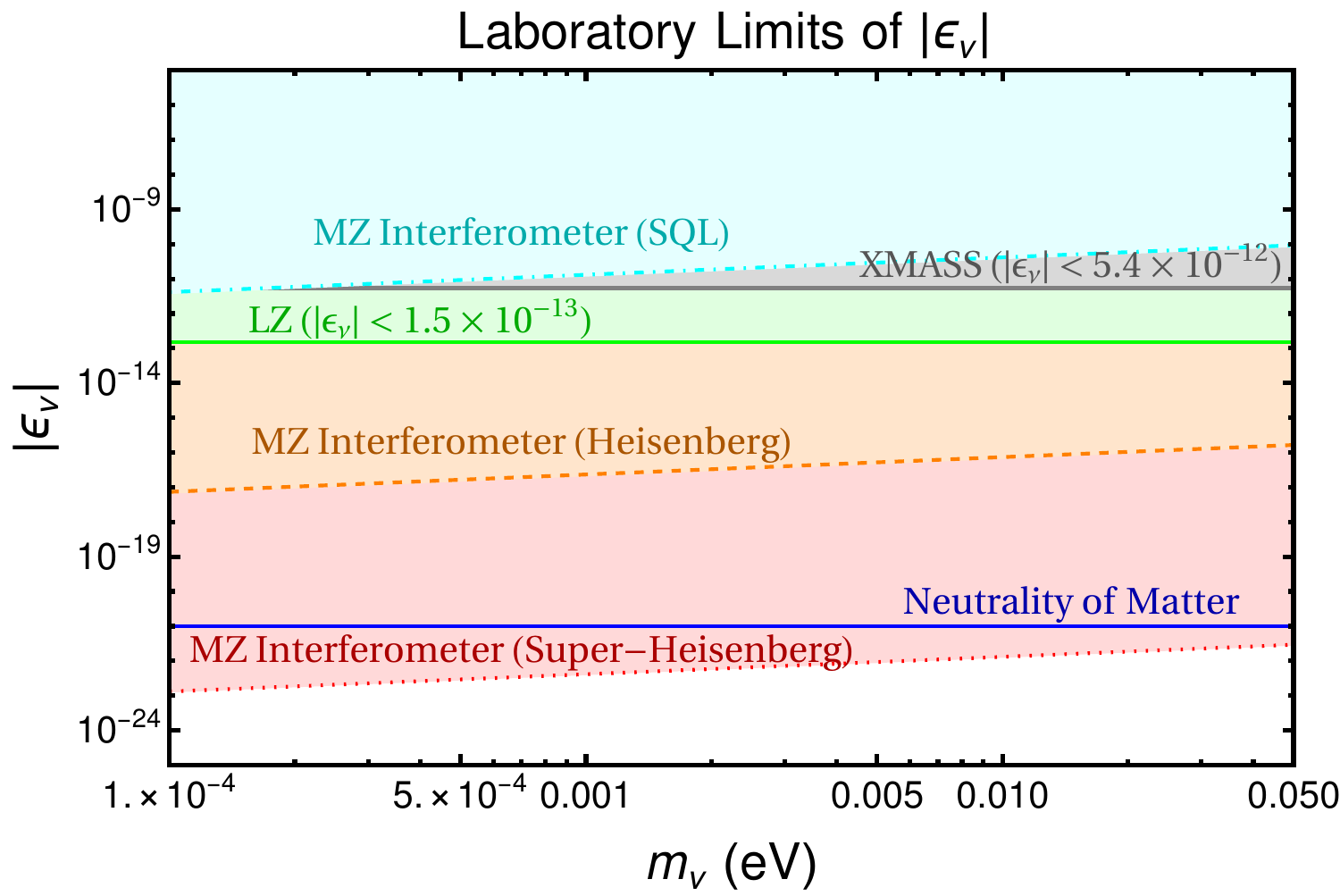} }}
	\caption{The sensitivity reach of the MZ interferometer with arm length L = 1 km and 1.17 eV laser for $|n_{\nu} - n_{\bar{\nu}}|/n_{\text{CNB}} = 10^{-8}$ vs the existing laboratory limits. Here, we take $n_{\text{CNB}} = 56\, \text{cm}^{-3}$ and set the phase sensing photon number $N_{ps} = 10^{23}$~\cite{GammeVT-969:2007pci,Bahre:2013ywa,ALPS:2009des,Inada:2013tx}. } 
	\label{fig:sensitivity_labs}
\end{figure}

It is clear from Eq. \eqref{eq:delta} that the CNB electric charge is evaluated at the mass basis since
it depends explicitly on the neutrino mass. The existing astrophysical bounds on the electric charge of
the neutrino in mass basis are shown as the light gray area and the light green are in Fig.\ref{fig:sensitivity_astro}. Both limits are extracted from in-medium photon decay to a pair of neutrino
anti-neutrino in the star. This leads to the cooling of the star with strong
electromagnetic field (magnetar) which sets $|\epsilon_{\nu}| < 10^{-12}$~\cite{Das:2020egb}. In addition, the same process induces the enhancement of the tip of the red
giant branch (TRGB) star brightness which gives more stringent limit~$|\epsilon_{\nu}| < 6.3 \times 10^{-15}$~\cite{Fung:2023euv}. For neutrino mass range  $10^{-4}\, \text{eV} \leq m_{\nu} \leq 0.05$ eV allowed by the particle data group (PDG)~\cite{ParticleDataGroup:2024cfk}, the projected
sensitivities of our proposal are shown in light cyan ($4.1 \times 10^{-12} \leq \epsilon_{\nu} \leq 9.3 \times 10^{-11}$), light orange ($7.4\times 10^{-18} \leq \epsilon_{\nu} \leq 1.6 \times 10^{-16}$), as well as light red ($1.3 \times 10^{-23} \leq \epsilon_{\nu} \leq 2.9 \times 10^{-22}$) region
corresponding to three different operating modes of the MZ interferometer: the SQL, the Heisenberg limit
and super-Heisenberg limit, respectively. As can been seen from Fig.~\ref{fig:sensitivity_astro}, the astrophysical bounds are stronger than the
projected limit from the SQL mode. It should be noted that the astrophysical limits are model dependent  and suffer from astrophysical uncertainties. On the other hand, when the MZ interferometer operates at both Heisenberg and
super-Heisenberg mode, it surpasses the astrophysical constraints. As an additional remark, there is more
stringent limit from cosmology $\epsilon < 4 \times 10^{-35}$ as cited by the
PDG~\cite{ParticleDataGroup:2024cfk}. This controversial bound~\cite{Karshenboim:2024iff,Giunti:2024gec}  is derived from the charge asymmetry of the universe under
model dependent assumptions: the infnite conductivity of the universe, the charge was
generated during primordial phase transition, and the
charge was distributed uniformly throughout the universe~\cite{Caprini:2003gz}.

On the other hand, several laboratory experiments have set the limits on the electric charge of the
neutrino. The XMASS collaboration puts the neutrino charge in the mass basis to be lower than $5.4 \times 10^{-12}$ at $90 \%$ confident level~\cite{XMASS:2020zke} shown as the light gray region in
Fig.~\ref{fig:sensitivity_labs}. Moreover, more stringent constraint on neutrino electric charge in the
same basis comes from LZ experiment $ |\epsilon_{\nu} | < 1.5 \times 10^{-13}$~\cite{AtzoriCorona:2022jeb} as displayed by the light green area in the same figure. Finally, the neutrality of
matter have placed the strongest bound on the neutrino charge to be about $10^{-21}$~\cite{Bressi:2011crt} (blue solid line). Clearly, the best projected sensitivity of our proposal would exceed the existing lab limits when it works at super-Heisenberg mode.   
    
\section{Summary and Conclusion}
\label{sec:Summary}

As one of the important pieces of the Big Bang theory, the detection of cosmic neutrino background plays an
important role in cosmology, astrophysics, as well as particle physics. Due to its elusive nature,
vanishing cross section and energy, the discovery of such relic poses a great challange to the
experimentalists. In addition, it has been shown that the CNB accumulates at the surface of the earth
with $10^{8}$ suppression factor of its corresponding number density $n_{\text{CNB}}$ in the present universe.

We propose to employ Mach-Zehnder interferometer to probe this terrestrial accumulation provided the
CNB carries non-zero electric charge. For neutrino mass equals to 0.05 eV and requiring the SNR to be greater than one, we demonstrate that our proposal
can probe the fractional electric charge of the CNB $\epsilon_{\nu}$
as low as: $9.3 \times 10^{-11}$, $1.6 \times 10^{-16}$, and $2.9 \times 10^{-22}$ provided the
inteferometer works under the SQL, the Heisenberg, and super-Heisenberg limit, respectively. This is
quite competitve with the existing bounds from both astrophysics and experiments. Moreover, the super
Heisenberg operational mode of the interferometer is shown to be better than the most stringent limit
from neutrality of matter. Thus, we show that our proposal offers a potential venue to detect the relic neutrino.

\section*{Acknowledgment}
\label{sec:Acknowledgment}
We would like to acknowledge the support of National Center for Theoretical Sciences (NCTS). This work was supported in part by the National Science and Technology Council (NSTC) of Taiwan under Grant No.NSTC 113-2112-M-003-007 and 113-2811-M-003-019.

\end{document}